\documentstyle[aps,epsfig]{revtex}
\twocolumn
\begin{document}
\twocolumn[\hsize\textwidth\columnwidth\hsize\csname@twocolumnfalse\endcsname
\title{Oscillatory disintegration of a trans-Alfv\'{e}nic shock: A 
magnetohydrodynamic simulation}
\author{S. A. Markovskii}
\address{Sternberg Astronomical Institute, Moscow State University, Moscow 
119899, Russia}
\author{S. L. Skorokhodov}
\address{Computer Center, Russian Academy of Sciences, Moscow 117967, Russia}
\maketitle
\begin{abstract}
Nonlinear evolution of a trans-Alfv\'{e}nic shock wave (TASW), at which the 
flow velocity passes over the Alfv\'{e}n velocity, is computed in a 
magnetohydrodynamic approximation. The analytical theory suggests that an 
infinitesimal perturbation of a TASW results in its disintegration, i.e., 
finite variation of the flow, or transformation into some other unsteady 
configuration. In the present paper, this result is confirmed by numerical 
simulations. It is shown that the disintegration time is close to its 
minimum value equal to the shock thickness divided by a relative velocity of 
the emerging secondary structures. The secondary TASW that appears after the 
disintegration is again unstable with respect to disintegration. When the 
perturbation has a cyclic nature, the TASW undergoes oscillatory 
disintegration, during which it repeatedly transforms into another TASW. This 
process manifests itself as a train of shock and rarefaction 
waves, which consecutively emerge at one edge of the train and merge at the 
other edge.
\end{abstract}
\vskip1pc]
\narrowtext

\section{Introduction}

It has long been believed that trans-Alfv\'{e}nic shock waves (TASWs), at 
which the flow velocity passes over the Alfv\'{e}n velocity, cannot exist in 
the real world. Since a stationary trans-Alfv\'{e}nic shock transition was 
obtained in a numerical simulation \cite{wu87}, this conventional view point 
was replaced by an opposite view point. The overall claim was that there is 
no principal difference between TASWs and fast and slow shocks, at which the 
flow is super- and sub-Alfv\'{e}nic, respectively. At the same time, the 
contradiction inherent in a stationary TASW, which follows from an 
analytical theory, was not lifted. To reconcile this contradiction, it was 
suggested that a TASW exists in an unsteady state in which it is repeatedly 
destroyed and recovered \cite{m98}. In the present paper, we show by way of 
magnetohydrodynamic (MHD) simulation that the evolution of a TASW may have 
the form of oscillatory disintegration, i.e., reversible transformation into 
another TASW.

The disintegration of an arbitrary hydrodynamic discontinuity was considered 
for the first time by Kotchine \cite{kotchine}. After that, Bethe 
\cite{bethe} studied the disintegration of shock waves. In the absence of a 
magnetic field, the shock may disintegrate only in a medium with anomalous 
thermodynamic properties. The magnetic field enlarges the number of possible 
discontinuous structures thus giving additional degrees of freedom for the 
disintegration. The disintegration configurations of arbitrary MHD 
discontinuities were obtained in Refs. \cite{arbit}. Furthermore, it has 
been shown that trans-Alfv\'{e}nic shock transitions can be realized also 
through a set of several discontinuities \cite{shock}, in contrast with fast 
and slow transitions. However, this fact on its own does not assure that the 
shock disintegrates. 

The important feature that predetermines the disintegration of TASWs is their 
nonevolutionarity. The problem of evolutionarity was initially formulated for 
the fronts of combustion \cite{landau} and hydrodynamic 
discontinuities \cite{cf}. Evolutionarity is a 
property of a discontinuous flow to evolve in such a way that the flow 
variation remains small under the action of a small perturbation. This is not 
the case for a nonevolutionary discontinuity. At such a discontinuity, the 
system of boundary conditions, which follow from the conservation laws, does 
not have the unique solution for the amplitudes of outgoing waves generated 
by given incident waves. From a mathematical view point this means that the 
number of unknown parameters (the amplitudes of the outgoing waves and the 
discontinuity displacement) is incompatible with the number of independent 
equations. Since a physical problem must have the unique solution, the 
assumption that the perturbation of a nonevolutionary discontinuity is 
infinitesimal leads to a contradiction. In fact, the infinitesimal 
perturbation results in disintegration, i.e., finite variation of the initial 
flow, or transformation into some other unsteady configuration. 

The evolutionarity requirement gives additional restrictions on the flow 
parameters at a shock, compared to the condition of the entropy increase. The 
restrictions appear because the direction of wave propagation (toward the 
discontinuity surface or away from it), and thus the number of the outgoing 
waves, depends on the flow velocity. If the velocity is large enough then the 
given wave may be carried down by the flow. Therefore, at an evolutionary 
discontinuity, the flow velocity must be such that it provides the 
compatibility of the boundary equations. This form of evolutionarity 
condition was applied to MHD shock waves in Refs. \cite{mhd,s58}.  As a 
result, the fast and slow shocks are evolutionary, while the 
TASWs are nonevolutionary. 

This classical picture was challenged when Wu \cite{wu87} obtained a 
stationary TASW in a numerical simulation. The existence of a stationary 
numerical solution does not mean of course that the shock is stable with 
respect to disintegration or transition into another unsteady flow. Wu 
\cite{wu88} demonstrated that a TASW, which is subfast upstream and subslow 
downstream, disintegrates under the action of a small Alfv\'{e}n perturbation 
with a large enough characteristic time.  Nevertheless, this numerical 
result was interpreted as being in a contradiction with the principle of 
evolutionarity and stimulated the efforts to modify or even disprove this 
principle.

It was suggested that the free parameters that describe a nonunique structure 
of a TASW \cite{wu90} or the amplitudes of strongly damping dissipative waves 
\cite{hada} should be included in the number of unknown parameters when 
solving the problem of evolutionarity. This would make the TASW evolutionary.  
In both cases, however, the perturbation is confined within the shock 
transition layer. Consequently, it does not enter into the boundary 
conditions, which relate the quantities far enough from the transition layer, 
and thus it does not contribute to the evolutionarity \cite{evol}.

Wu \cite{wu90} also argued that the TASW whose nonevolutionarity is 
based on separation of Alfv\'{e}n perturbations from the remaining 
perturbations \cite{s58} becomes evolutionary in the case of a nonplanar shock 
structure because in this case the separation formally does not take place. 
However, as shown by Markovskii \cite{evol}, the coupling of the 
small-amplitude Alfv\'{e}n modes with a low enough frequency to the remaining 
modes is weak (unless the shock is of the type close to one of the degenerate 
types, Alfv\'{e}n discontinuity or switch shocks). Therefore the coupling 
becomes essential only when the small perturbation generates large variation 
of the flow, which is the same result as predicted by the principle of 
evolutionarity.

There is one more finding that favors the nonexistence of stationary 
TASWs. As discussed by Kantrowitz and Petschek \cite{kp}, the TASWs are 
isolated solutions of Rankine-Hugoniot problem, which do not have neighboring 
solutions corresponding to small deviations of boundary conditions. Wu and 
Kennel \cite{wk} introduced a new class of trans-Alfv\'{e}nic 
shock-like structures with noncoplanar boundary states. The thickness of 
such a structure increases in the course of time, and it eventually evolves 
to a large-amplitude Alfv\'{e}n wave. It was thus shown that neighboring to a 
TASW are time-dependent configurations, which are not solutions of the 
Rankine-Hugoniot problem. In addition, Falle and Komissarov \cite{fk} 
recently considered stationary TASWs of all possible types and showed that 
the shocks disintegrate if the boundary values deviate from their initial 
values.

Strictly speaking, a TASW, in contrast with fast and slow shocks, becomes a 
time-dependent shock-like structure once it is perturbed by a small-amplitude 
Alfv\'{e}n wave because the Alfv\'{e}n wave violates the coplanarity 
condition. This fact, already on its own, means that the TASW becomes 
unsteady under the action of the small perturbation. However, the scenario 
for its evolution depends on the initial configuration and on the nature of 
the perturbation. After the disintegration, the magnetic field reversal given 
at the initial nonevolutionary shock may be taken either by a secondary TASW 
or by an Alfv\'{e}n discontinuity. Both structures are nonevolutionary 
\cite{m98,evol}. Therefore single disintegration does not lift the 
contradiction inherent in a TASW. The main question that we solve in this 
paper is what happens to the post-disintegration nonevolutionary 
configuration. We show that the secondary TASW is again unstable with respect 
to disintegration and that the evolution of a TASW may have the form of 
oscillatory disintegration.  In Sec.  II, we describe the simulation method. 
In Sec. III, we discuss the results of the calculations. Our conclusions are 
presented in Sec. IV.

\section{Numerical method}

We take the MHD equations in the following form
\begin{mathletters}
\label{1}
\begin{equation}
{\partial \rho \over \partial t} + {\partial \rho v_{x} \over \partial x}  
= 0,
\label{1a}
\end{equation}
\begin{equation}
{\partial \rho   v_{x} \over \partial t} + {\partial \over \partial x} 
\biggl(  p + \rho  v_{x}^{2}  + {\textstyle {1\over 2}}{\bf B}_{\perp}^{2}  - 
{\textstyle { 4\over 3}} \eta  {\partial v_{x} \over \partial x}\biggr) 
= 0,
\label{1b}
\end{equation}
\begin{equation}
{\partial \rho  {\bf v}_{\perp} \over \partial t} + {\partial \over 
\partial x}   \biggl(  \rho  v_{x}  {\bf v}_{\perp}  - B_{x} {\bf B}_{\perp} 
- \eta {\partial {\bf v}_{\perp} \over \partial x} \biggr) = 0,
\label{1c}
\end{equation}
\begin{equation}
{\partial {\bf B}_{\perp} \over \partial t} + {\partial \over \partial 
x} \biggl( v_{x} {\bf B}_{\perp}  - B_{x} {\bf v}_{\perp} 
- \nu_{m} {\partial {\bf B}_{\perp} \over \partial x} \biggr) = 0,
\label{1d}
\end{equation}
\begin{eqnarray}
&& {\partial \over \partial t}  \biggl( {\textstyle {1\over 2}} \rho v^{2}  
+ { p \over \gamma - 1 } + {\textstyle {1\over 2}} {\bf B}_{\perp}^{2} 
\biggr) + {\partial \over \partial x}   \biggl[   \rho  v_{x}  
\biggl({\textstyle {1\over 2}} v^{2} 
\nonumber \\
&& \quad + { \gamma 
\over \gamma - 1}  { p\over \rho} \biggr) 
+ \biggl( {\bf B}_{\perp} \cdot \biggl( v_{x} {\bf B}_{\perp}  - B_{x} {\bf 
v}_{\perp} 
- \nu_{m} {\partial {\bf B}_{\perp} \over \partial x} \biggr) 
\biggr) 
\nonumber \\
&& \quad - \eta \biggl( {\textstyle{4\over3} }
v_{x} {\partial v_{x}\over \partial x} 
+ \biggl( {\bf v}_{\perp} \cdot 
{\partial {\bf v}_{\perp}\over \partial x} 
\biggr)\biggr)\biggr] = 0.
\label{1e}
\end{eqnarray}
\end{mathletters}
Here the subscript "$\perp$" denotes the vector component perpendicular to 
the $x$ axis, $B_{x}={\rm const},$ magnetic diffusivity $\nu_{m}$ and 
viscosity $\eta$ are put constant and equal to 0.1 in all calculations, and 
we use the units such that the factor $4\pi$ does not appear. The initial 
distribution of the MHD quantities is given by the following formulas
\begin{mathletters}
\label{2}
\begin{equation}
\rho = { \textstyle {1\over 2}} (\rho_{\uparrow} + \rho_{\downarrow}) - 
{ \textstyle {1\over 2}} (\rho_{\uparrow} - \rho_{\downarrow}) {\rm tanh} 
( { x/ L}) ,
\label{2a}
\end{equation}
\begin{equation}
v_{x} = { \textstyle {1\over 2}} (v_{x\uparrow} + v_{x\downarrow}) - 
{\textstyle {1\over 2}} (v_{x\uparrow} - v_{x\downarrow}) {\rm tanh} ( { 
x/ L} ) ,
\label{2b}
\end{equation}
\begin{equation}
p = {\textstyle {1\over 2}} (p_{\uparrow} + p_{\downarrow}) - 
{ \textstyle {1\over 2}} (p_{\uparrow} - p_{\downarrow}) {\rm tanh} ( { 
x/ L}) ,
\label{2c}
\end{equation}
\begin{equation}
B_{y} = B_{\tau} {\rm cos} (\theta), \qquad
B_{z} = B_{\tau} {\rm sin} (\theta),
\label{2d}
\end{equation}
\begin{equation}
v_{y} = v_{\tau} {\rm cos} (\theta), \qquad
v_{z} = v_{\tau} {\rm sin} (\theta),
\label{2e}
\end{equation}
\begin{eqnarray}
B_{\tau} = && {\textstyle {1\over 2}} (\mid B_{\perp\uparrow}\mid + \mid 
B_{\perp\downarrow}\mid ) 
\nonumber \\
&& - {\textstyle {1\over 2}} 
(\mid B_{\perp\uparrow} \mid 
- \mid B_{\perp\downarrow}\mid ) {\rm tanh} ( { x/ L} ),
\label{2f}
\end{eqnarray}
\begin{eqnarray}
v_{\tau} = && {\textstyle {1\over 2}} (\mid v_{\perp\uparrow}\mid  + \mid 
v_{\perp\downarrow}\mid ) 
\nonumber \\
&& - {\textstyle {1\over 2} }
(\mid v_{\perp\uparrow} \mid - \mid 
v_{\perp\downarrow}\mid ) {\rm tanh} ( { x/ L} ), 
\label{2g}
\end{eqnarray} 
\begin{equation} 
\theta = {\textstyle {\pi\over 2}} (1 + {\rm tanh} ( { x/ L})),
\label{2h}
\end{equation}
\end{mathletters}
where the subscripts "$\uparrow$" and "$\downarrow$" denote the quantities in 
the asymptotic upstream and downstream regions, respectively. 

After the 
configuration relaxes to a steady state, it is perturbed by an Alfv\'{e}n 
wave specified by the expression
\begin{mathletters}
\label{3}
\begin{equation}
B_{z} = {\textstyle {1 \over 2}} \delta B_{z} \biggl( 1 + {\rm tanh} \biggl( 
{x - x_{0} \over l} \biggr) \biggr) ,
\label{3a}
\end{equation}
\begin{equation}
v_{z} = - B_{z} / \sqrt{\rho}.
\label{3b}
\end{equation}
\end{mathletters}
This wave moves downstream. The configuration is set by putting 
$B_{x}=0.89,$ $L=1.45,$ and
\begin{mathletters}
\label{4}
\begin{equation}
B_{y\uparrow} = 0.93, \qquad B_{y\downarrow} = -0.8,
\label{4a}
\end{equation}
\begin{equation}
B_{z\uparrow} = v_{z\uparrow} = 0., \qquad 
B_{z\downarrow} = v_{z\downarrow} = 0.,
\label{4b}
\end{equation}
\begin{equation}
v_{x\uparrow} = 1., \qquad v_{x\downarrow} = 0.55042,
\label{4c}
\end{equation}
\begin{equation}
v_{y\uparrow} = 1.04494, \qquad v_{y\downarrow} = -0.49476,
\label{4d}
\end{equation}
\begin{equation}
\rho_{\uparrow} = 1., \qquad \rho_{\downarrow} = 1.81681,
\label{4e}
\end{equation}
\begin{equation}
p_{\uparrow} = 0.00116, \qquad p_{\downarrow} = 0.56319.
\label{4f}
\end{equation}
\end{mathletters}
This corresponds to a ${\rm II \rightarrow III}$ 
shock, for which $V_{+\uparrow} > 
v_{x\uparrow} > V_{Ax\uparrow}$ and $V_{Ax\downarrow} > v_{x\downarrow} > 
V_{-\downarrow},$ where $V_{+}$ and $V_{-}$ are the fast 
and slow magnetosonic velocities.

We solve Eq. (\ref{1}) using a uniform grid and an explicit conservative 
Lax-Wendroff finite-difference scheme with physical dissipation \cite{pt}. 
The time step is limited by the Courant-Friedrichs-Lewy (CFL) condition and 
by the dissipation timescale. The boundary values are obtained by hyperbolic 
interpolation. The numerical interval $-50 < x < +300$ is covered by 2600 
grid points. The interval is chosen in such a way that no 
large-amplitude wave reaches the boundaries during the computation time. 
Small-amplitude waves pass through the boundaries without any detectable 
reflection which could affect the flow inside the simulation region. We have 
tested our code for a smaller mesh and a corresponding time step determined 
by the CFL condition as well as for the same mesh and a time step smaller 
than that determined by the CFL condition. The test showed that there is no 
considerable dependence of our results on the mesh size and 
time step.

\section{Results of simulations}

\begin{figure*}
\begin{center}
\epsfig{file=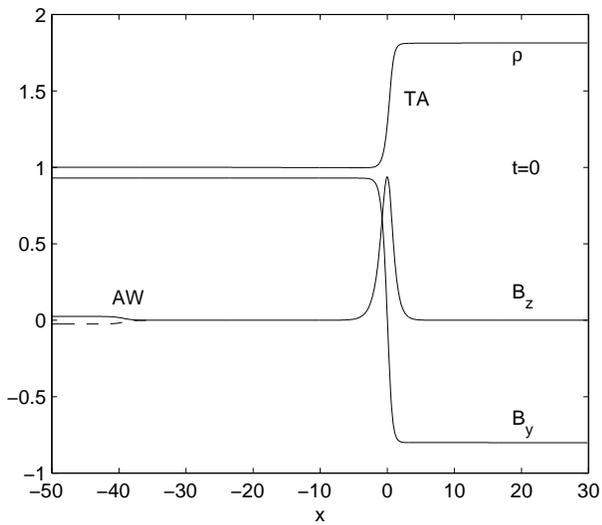,width=8cm} 
\end{center}
\caption{Inintial distribution of the density and magnetic field at the 
TASW perturbed by a small-amplitude Alfv\'{e}n wave with positive (solid 
line) and negative (dashed line) value of $ \delta B_{z}$.} 
\label{f1} 
\end{figure*}

\begin{figure*}
\begin{center}
\epsfig{file=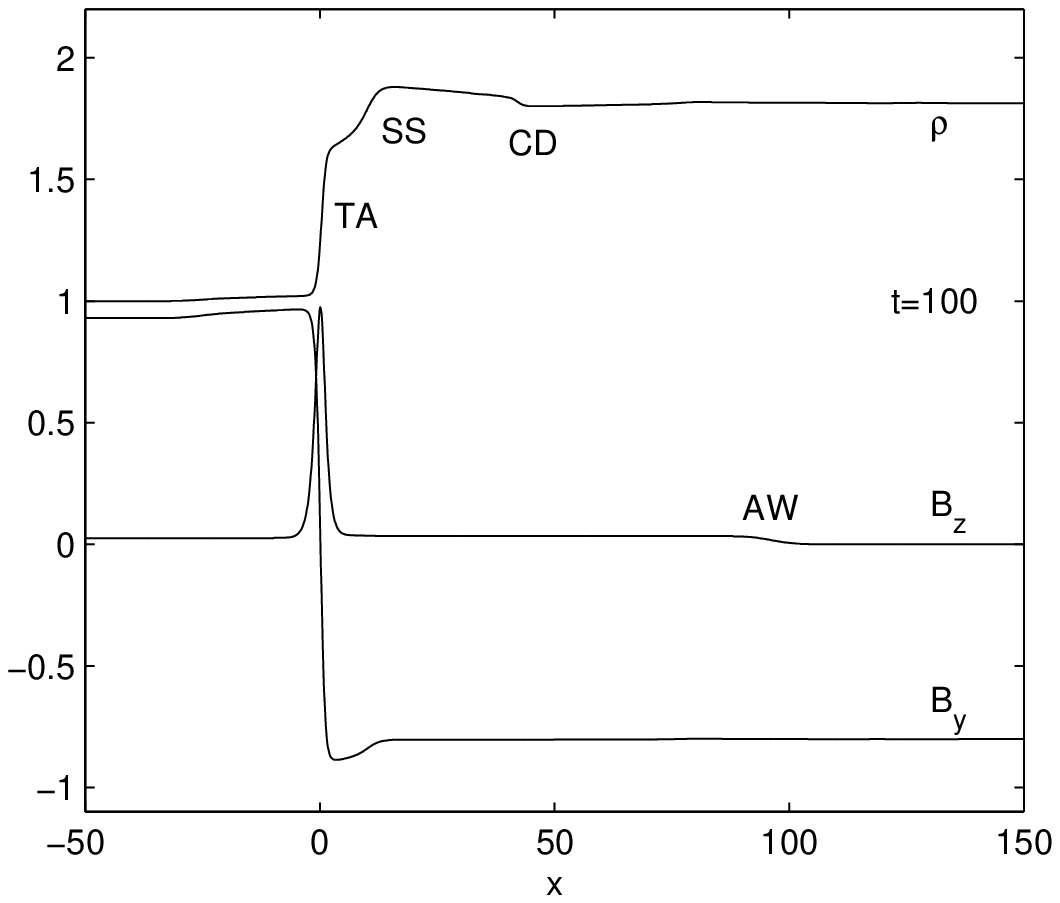,width=8cm} 
(a)
\epsfig{file=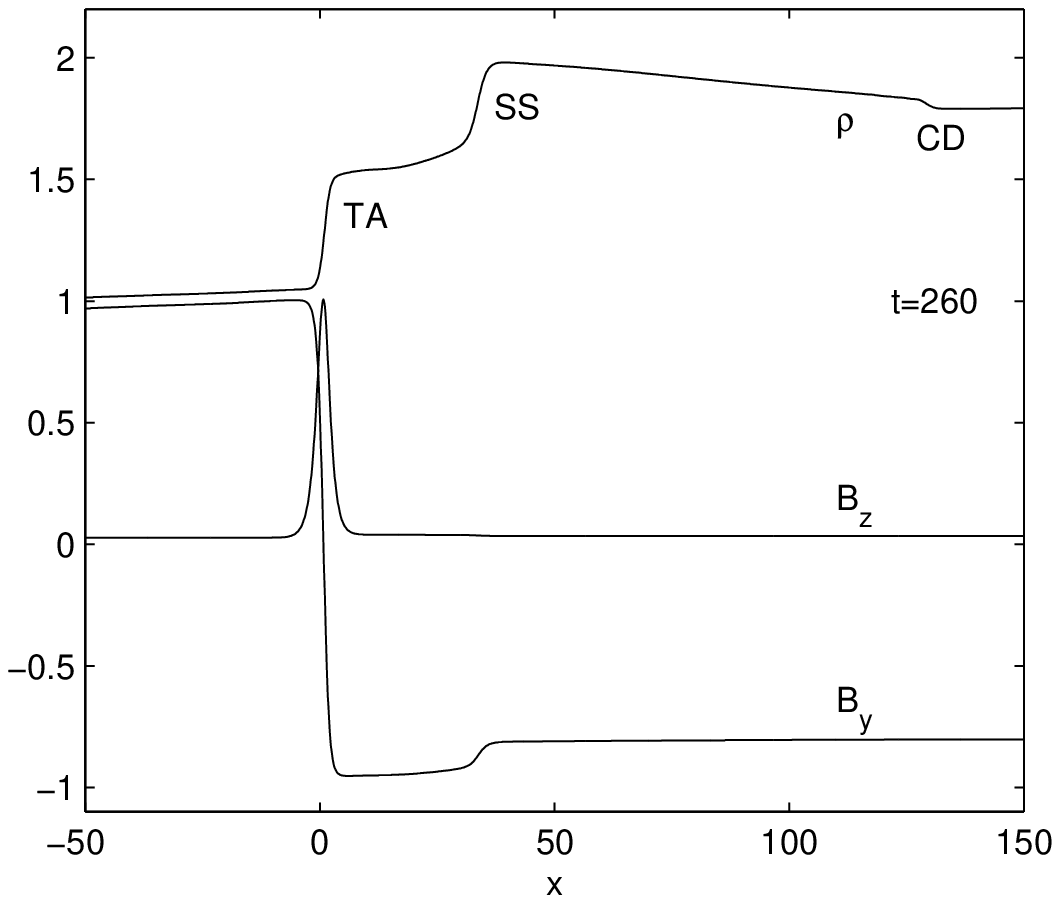,width=8cm} 
(b)
\epsfig{file=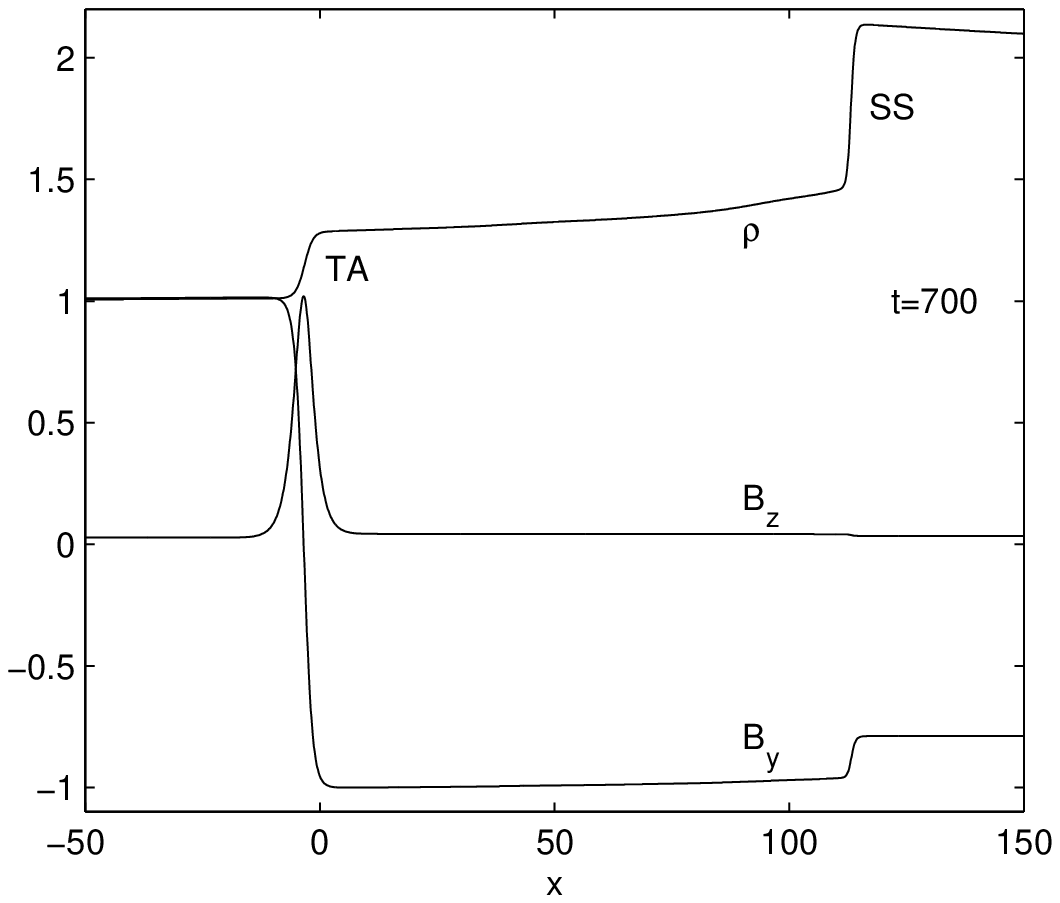,width=8cm} 
(c)
\end{center}
\caption{Disintegration configuration for a constant positive perturbation 
at $t=100$ (a), $t=260$ (b), and $t=700$ (c).} 
\label{f2} 
\end{figure*}

Equation (\ref{2}) does not exactly 
describe the shock structure. Therefore the flow undergoes time variations 
until it adjusts to a stationary shock transition. The resulting boundary 
values differ slightly from those given by 
Eq. (\ref{4}) but the difference is less than 1\%. The conservation laws for 
these new values are fulfilled with the precision less than 0.1\%. The 
stationary configuration is then perturbed by a small-amplitude Alfv\'{e}n 
wave with  $l=-L,$ $x_{0}= -40,$ and $\delta B_{z}= 0.025$ or $\delta B_{z}= 
-0.025$ (Fig.~\ref{f1}). Note that $\delta B_{z}$ is about 50 times smaller 
than $\mid {\bf B}_{\uparrow} \mid.$ Although in the case of an upstream 
incident wave the perturbation of $B_{z}$ and $v_{z}$ (not shown) is carried 
to the downstream region, the boundary conditions for the Alfv\'{e}n waves 
are incompatible. Therefore the given Alfv\'{e}n perturbation pumps $B_{z}$ 
and $v_{z}$ into the shock or out of the 
shock, depending on the sign of $B_{z}$ inside the transition layer. Since 
$B_{z}$ inside the transition layer is nonzero, the shock behaves in 
different ways under the action of the perturbations with positive and 
negative $\delta B_{z}.$ If the shock and the perturbation carry $B_{z}$ of 
the same sign, the shock disintegrates into a ${\rm II \rightarrow III}$ 
shock of a smaller amplitude, a large-amplitude slow shock, 
and some other structures of a much smaller amplitude (Fig.~\ref{f2}a,b). 

\begin{figure*}
\begin{center}
\epsfig{file=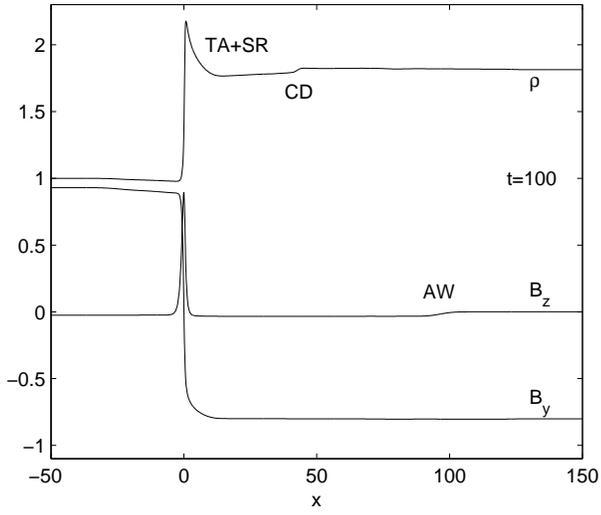,width=8cm} 
(a)
\epsfig{file=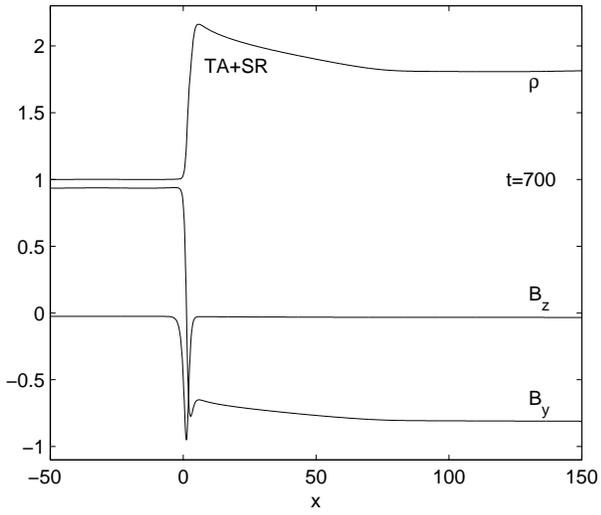,width=8cm} 
(b)
\epsfig{file=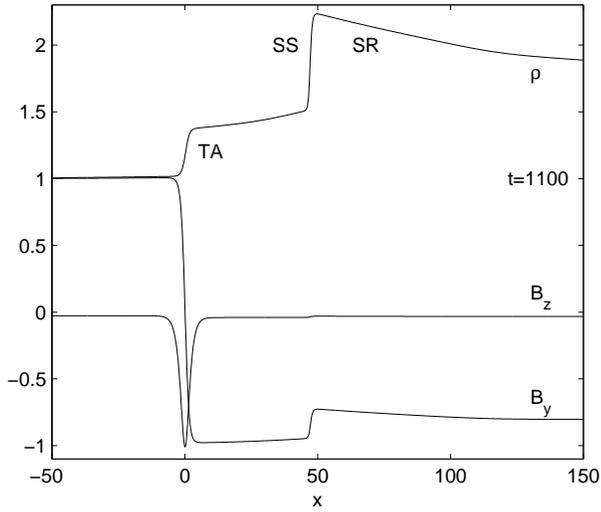,width=8cm} 
(c)
\end{center}
\caption{Disintegration configuration for a constant negative perturbation 
at $t=100$ (a), $t=700$ (b), and $t=1100$ (c).} 
\label{f3} 
\end{figure*}

If the shock and the perturbation carry $B_{z}$ of opposite signs, the 
situation is somewhat peculiar. The main secondary structures are a TASW and 
a slow rarefaction (Fig.~\ref{f3}a,b). However, these structures do not 
become separated. The reason is that the secondary TASW is of a so-called 
${\rm II \rightarrow IV=III}$ type \cite{kbw}. This means that the downstream 
velocity at the shock is exactly equal to the slow magnetosonic velocity. 
Therefore there is no disintegration in the usual sense but the 
configuration becomes unsteady because the right boundary of the slow 
rarefaction moves away from the TASW, while the left 
boundary remains attached to the shock. Note that the rarefaction wave is 
attached to the TASW not at the density peak but somewhere to the right of 
the peak. This is related to the fact that the density profile of a ${\rm II 
\rightarrow IV}$ shock has a maximum (see, e.g., Ref. \cite{wu88}), in 
contrast with the monotonic profile of a ${\rm II \rightarrow III}$ shock.

From the moment when the Alfv\'{e}n wave with $\delta B_{z}>0$
arrives to the shock, the disintegration starts almost immediately, in 
contrast with the result of Wu \cite{wu88}. The reason is that the 
disintegration time depends on the shock type and on its initial state. This 
can be understood as follows. The important characteristic of a TASW, 
introduced by Kennel {\it et al.} \cite{kbw}, is the integral of $B_{z}$ over 
the 
transition layer, 
\begin{equation} 
I_{z} = \int \limits_{x\downarrow}^{x\uparrow} B_{z} dx.  
\label{5} 
\end{equation} 
This integral fixes the nonunique structure of a TASW. For a ${\rm II 
\rightarrow III}$ shock, the quantity $I_{z}$ takes two distinct values, 
$I_{z0}$ and $-I_{z0},$ and, for a ${\rm I \rightarrow III}$ or ${\rm II 
\rightarrow IV}$ shock, it falls into the interval $-I_{z0} < I_{z} < 
I_{z0}.$ The quantity $I_{z0}$ depends on the boundary values, and it tends 
to infinity when the shock approaches an Alfv\'{e}n discontinuity or a switch 
shock, which is intermediate between evolutionary and nonevolutionary shocks. 
This result was obtained for almost parallel small-amplitude shocks, but one 
may expect that it remains qualitatively valid in the general case.

When an Alfv\'{e}n wave is incident on a TASW, it changes $I_{z}.$ If we 
start from a planar ${\rm I \rightarrow III}$ or ${\rm II \rightarrow IV}$ 
shock ($I_{z}=0$), as in the case studied by Wu \cite{wu88}, the quantity 
$\mid I_{z} \mid $ first has to reach the value $I_{z0}.$ Only after that 
it falls into the forbidden region, and the disintegration starts. In the 
case of a ${\rm II \rightarrow III}$ shock, there is a different situation. 
Since $I_{z}$ takes only distinct values $I_{z0}$ and $-I_{z0},$ the 
disintegration starts immediately, and the disintegration time is close to 
its minimum value $L/V,$ approximately equal to 30 in our case, where $V$ is 
a relative velocity of the secondary discontinuities.

Let us now follow the further evolution of the post-disintegration 
configuration under the action of a small perturbation. Our main conclusion 
is that the secondary TASW is again unstable with respect to disintegration. 
At the same time, the way of evolution depends on a form of the 
perturbation. We first discuss the case where the perturbation of the 
secondary TASW is such that $I_{z}$ continues to increase or 
decrease, in particular where the perturbation is equal to its initial 
positive (Fig.~\ref{f2}b,c) or negative (Fig.~\ref{f3}b,c) value. If the 
perturbation of $B_{z}$ is positive then the shock spreads in space, with all 
the jumps, except for $\Delta B_{y}$ and $\Delta v_{y},$ decreasing in time. 
It thus approaches a large-amplitude Alfv\'{e}n wave. If the perturbation is 
negative, the shock first passes through the state in which $I_{z}=0.$ 
This is not in a contradiction with the analytical theory, because a ${\rm II 
\rightarrow IV}$ shock may have a planar structure, in contrast with a ${\rm 
II \rightarrow III}$ shock \cite{kbw}. When $\mid I_{z} \mid$ reaches a 
critical value, the shock disintegrates (Fig.~\ref{f3}b), and after that it 
spreads in space approaching a large-amplitude Alfv\'{e}n wave 
(Fig.~\ref{f3}c). The precursor of the disintegration is the peak in $B_{y}$ 
curve in Fig. \ref{3}b.

\begin{figure*}
\begin{center}
\epsfig{file=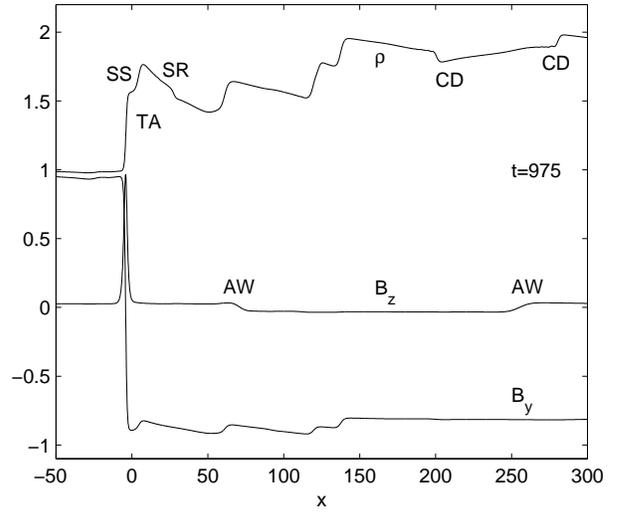,width=8cm} 
(a)
\epsfig{file=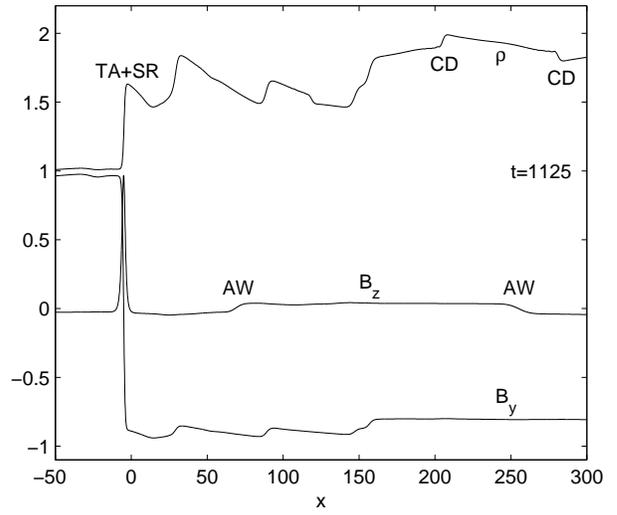,width=8cm} 
(b)
\epsfig{file=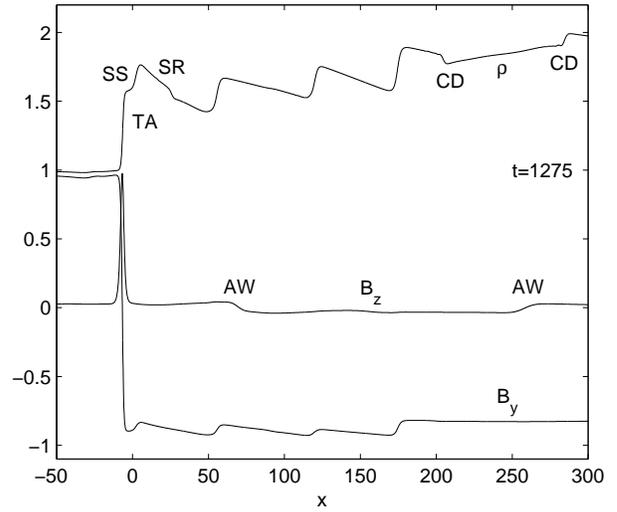,width=8cm} 
(c)
\end{center}
\caption{Configuration after several cycles of duration 150 time units at 
$t=975$ (a), $t=1125$ (b), and $t=1275$ (c). Oscillatory disintegration is 
started by a positive initial perturbation. The perturbation changes sign for 
the first time at $t=150.$ } 
\label{f4} 
\end{figure*}

\begin{figure*}
\begin{center}
\epsfig{file=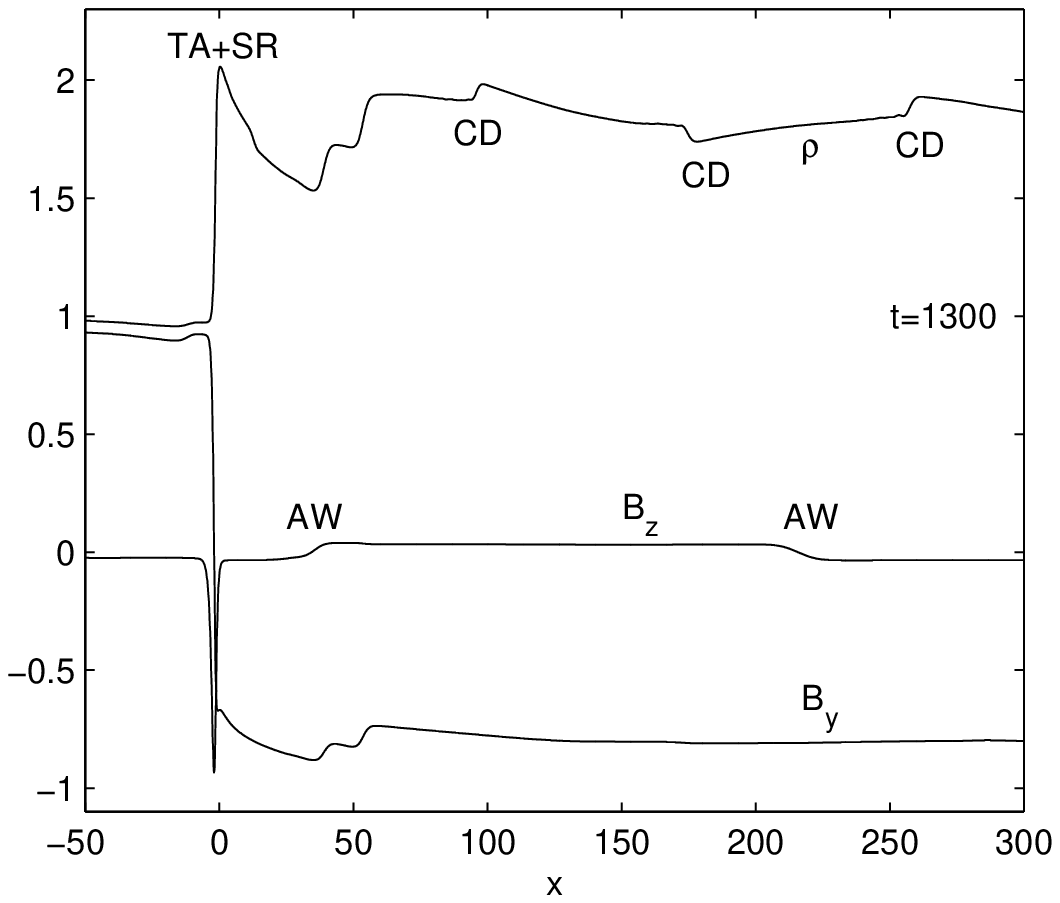,width=8cm} 
(a)
\epsfig{file=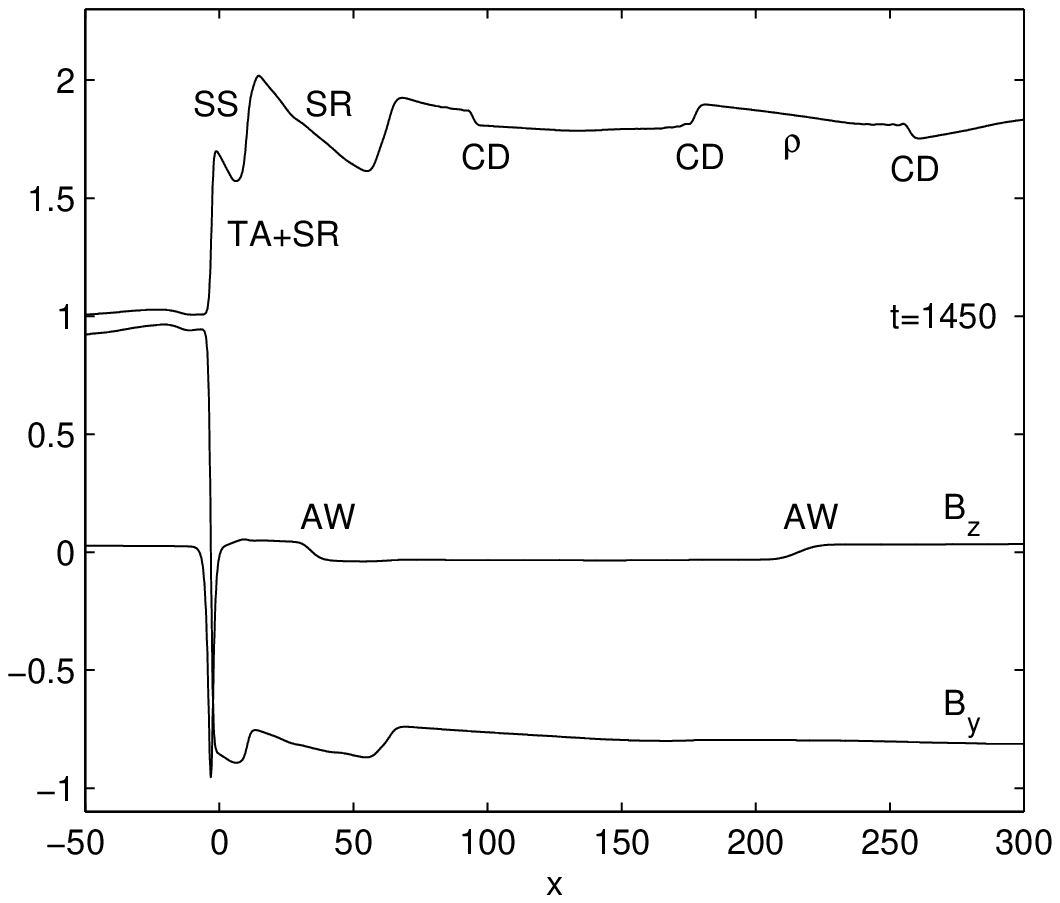,width=8cm} 
(b)
\epsfig{file=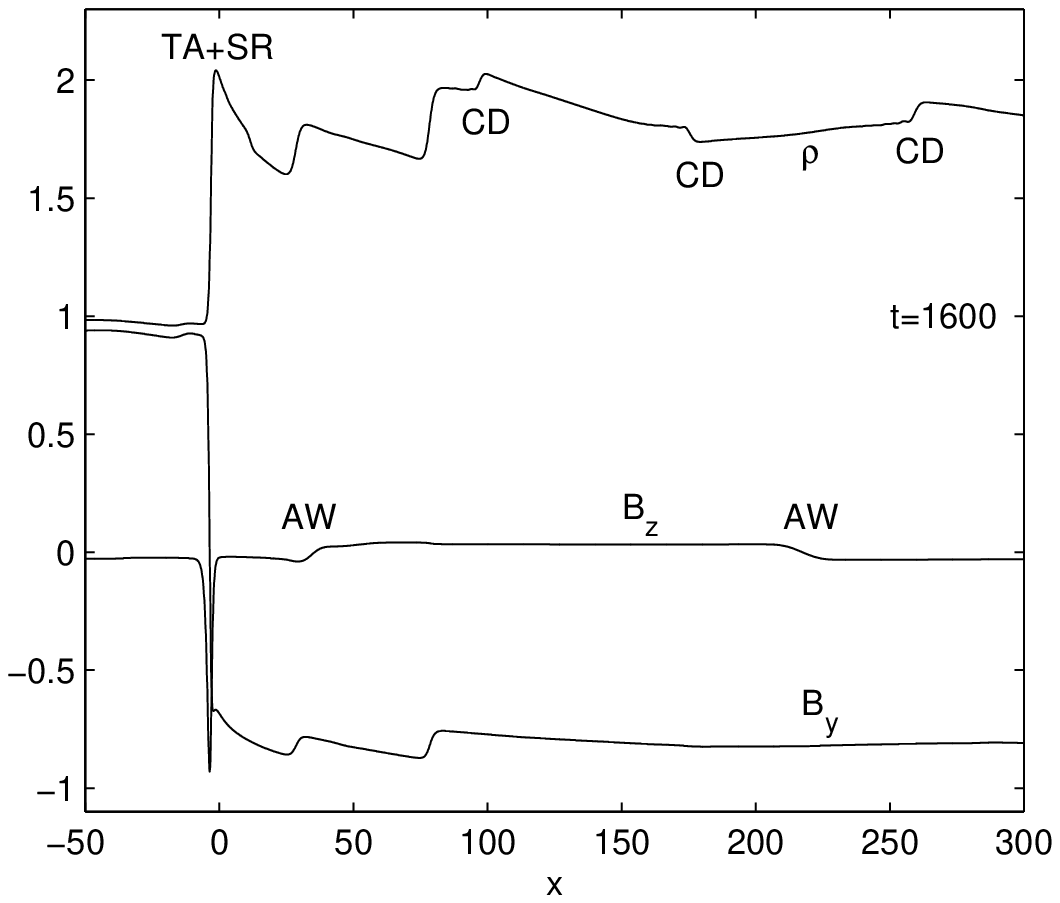,width=8cm} 
(c)
\end{center}
\caption{ Configuration after several cycles of duration 150 time units at 
$t=1300$ (a), $t=1450$ (b), and $t=1600$ (c). Oscillatory disintegration is 
started by a negative initial perturbation. The perturbation changes sign for 
the first time at $t=800.$ }
\label{f5} 
\end{figure*}

We now turn to a cyclic perturbation. We impose the perturbation described by 
Eq. (\ref{3}) in such a way that $B_{z}$ changes sign at $x=-40$ and the 
Alfv\'{e}n wave now carries the perturbation of the same amplitude but 
opposite sign. After the first disintegration starts (at $t=20$ for 
$\delta B_{z} >0$ and at $t=670$ for $\delta B_{z}<0$), the opposite sign 
perturbation arrives to the TASW each 150 units of time.  The resulting 
configuration is such that the increase of $\mid I_{z} \mid$ is repeatedly 
replaced by its decrease, and the shock undergoes oscillatory disintegration. 
The disintegration configurations after several cycles are shown in 
Figs.~\ref{f4} and \ref{f5}.  As can be seen from the figures, the 
configuration emits a sequence of contact discontinuities.  The contact 
discontinuities move with the flow velocity, which is approximately equal to 
that given by Eq.  (\ref{4c}). The corresponding time interval between the 
discontinuities is equal to 150. 

Downstream of the TASW, there is a wave train, which consists of slow shock 
and rarefaction waves. These structures are not standing in the flow. They 
consecutively emerge at the left edge of the train and merge at the right 
edge. The merging is seen in Fig. \ref{4}b at $x\approx 150.$ We note that, 
in the case of a negative initial perturbation, the transition through the 
state with $I_{z}=0$ is not necessary for the oscillatory 
disintegration to occur. If the perturbation changes sign for the first time 
before $I_{z}$ becomes negative, the disintegration configuration is similar 
to that shown in Fig.  \ref{5}, except for the sign of $I_{z}$ inside the 
shock.

Finally, the shock comes to a steady state only in a degenerate case where 
the perturbation of the secondary TASW exactly compensates the nonzero value 
of $B_{z}$ and $v_{z}$ outside of the transition layer. We emphasize that in 
all but the degenerate cases the small Alfv\'{e}n perturbation makes the TASW 
unsteady, in contrast with fast and slow shocks. However, there remains a 
question. Formally, the initial TASW becomes a time-dependent structure, much 
like the secondary TASW, since the Alfv\'{e}n perturbation arrives to the 
initial shock. The question is why the initial TASW disintegrates when 
$I_{z}$ increases monotonically, while the secondary TASW does not. To 
answer this question, we first mention that the secondary TASW is more 
close to a finite-amplitude Alfv\'{e}n wave than the initial shock. 
Alfv\'{e}n waves, as well as switch shocks, are singular structures. As shown 
by Kennel {\it et al.} \cite{kbw}, the quantity $dI_{z0}/dq$ tends to 
infinity as the shock approaches these singular structures. Here 
$q=B_{y\uparrow}/B_{y\downarrow}$ characterizes the jumps of the boundary 
values at the shock with a given $I_{z},$ and $I_{z0}(q)$ is an allowed 
curve in which a ${\rm II \rightarrow III}$ shock has a stationary structure.

Assume now that the initial shock is in the state $I_{z}=I_{z0}(q_{0}).$ A 
small Alfv\'{e}n perturbation changes $I_{z}.$ For the shock to remain in the 
curve $I_{z0}(q),$ a change of $q$ is required. In the general case, the 
variation of $I_{z}$ is comparable with the variation of $q,$ and thus with 
the jumps of the boundary values at the TASW. In this case, the evolution has 
the form of disintegration. By contrast, if the shock is close to the 
singular structure, the given variation of $I_{z}$ requires a small variation 
of $q,$ and the jumps of the boundary values are adjusted to $I_{z0}(q)$ in a 
diffusion-like manner. It should be mentioned that the curves $I_{z0}(q)$ 
were obtained by Kennel {\it et al.} \cite{kbw} for small-amplitude shocks 
propagating almost parallel to the magnetic field. Nevertheless, we speculate 
that, in our simulation, the initial TASW has a small enough 
$dI_{z0}/dq$ to disintegrate, while for the secondary TASW the quantity 
$dI_{z0}/dq$ is large enough to dim the disintegration. Such an explanation 
does not imply that a TASW cannot disintegrate more than one time in 
principle. Furthermore, in our simulation run with a positive constant 
$\delta B_{z},$ there is an evidence for a possible second disintegration at 
$t=380.$ However, the second disintegration is too faint to contend that it 
indeed takes place.

\section{Conclusions}

We have performed a numerical simulation of a trans-Alfv\'{e}nic shock wave.  
The shock that we have considered is of a ${\rm II \rightarrow III}$ type, 
i.e., it is subfast upstream and superslow downstream. We have shown that the 
shock disintegrates under the action of a small Alfv\'{e}n perturbation. The 
resulting configuration includes a secondary TASW, a large-amplitude slow 
shock or rarefaction wave, and other small-amplitude structures. We have also 
demonstrated that the secondary TASW is again unstable with respect to 
disintegration. When the perturbation has a cyclic nature, the shock 
undergoes an oscillatory disintegration. This result is in a qualitative 
agreement with our previous finding \cite{m98}. This process shows up 
as a train of slow shock and rarefaction waves, which 
consecutively emerge at one edge of the train and merge at the other edge. At 
the same time, the disintegration configuration of a small-amplitude almost 
parallel TASW discussed by Markovskii \cite{m98} includes alternating TASWs 
and Alfv\'{e}n discontinuities rather than alternating TASWs. This 
discrepancy is explained by the fact that, in the approximation used in Ref. 
\cite{m98}, the difference between the secondary TASW and the Alfv\'{e}n 
discontinuity manifests itself in higher orders.

In contrast with the results of Wu \cite{wu88}, the disintegration starts 
almost immediately after the Alfv\'{e}n perturbation arrives to the 
initial shock. The characteristic time of this process is equal to that 
required for the secondary structures to become separated. The reason for 
this can be seen as follows. TASWs have a nonunique structure. A ${\rm II 
\rightarrow IV}$ shock transition studied by Wu \cite{wu88}, as well as a 
${\rm I \rightarrow III}$ transition, allows a continuous family of integral 
curves, while the ${\rm II \rightarrow III}$ shock has two distinct integral 
curves. For given boundary values, each integral curve is fixed by the 
definite parameter. The incident Alfv\'{e}n wave changes the parameter and 
thus the shock structure. In the case of a ${\rm I \rightarrow III}$ or ${\rm 
II \rightarrow IV}$ shock, some time passes until the parameter falls into a 
forbidden region, and only after that the shock disintegrates. In the case of 
a ${\rm II \rightarrow III}$ shock, its structure immediately becomes 
inconsistent with the boundary values under the action of the Alfv\'{e}n 
wave, which initiates the disintegration. 

Thus, our simulations confirm that a TASW becomes unsteady when it is 
perturbed by a small-amplitude incident wave. Furthermore, an almost 
vanishing perturbation results in considerable dynamics at relatively small 
timescales. The scenario for the shock evolution depends on its initial state 
and on the nature of the perturbation.  In particular, the evolution may have 
the form of oscillatory disintegration in which the shock repeatedly 
transforms into another TASW.

\acknowledgments{This work is supported in part by Russian Foundation for 
Basic Research (grants 99-02-16344 and 98-01-00501).}

\end{document}